\documentclass[fleqn,twoside]{article}
\usepackage{espcrc2}
\usepackage{graphicx}
\usepackage[figuresright]{rotating}
\usepackage{lineno}

\newcommand{\ccpip}{\mathrm{CC1}\pi^+}
\newcommand{\ncpi}{\mathrm{NC}\pi^0}
\newcommand{\ccpipp}{\nu_{\mu}p\rightarrow\mu^-p\pi^+}

\newcommand{\ccpipcoh}{\nu_{\mu}A\rightarrow\mu^-A\pi^+}

\newcommand{\numu}{\nu_{\mu}}
\newcommand{\numubar}{\bar{\nu}_{\mu}}
\newcommand{\nue}{\nu_e}
\newcommand{\nuebar}{\bar{\nu}_e}
\newcommand{\mutoe}{\numu\rightarrow\nue}
\newcommand{\mubartoebar}{\numubar\rightarrow\nuebar}
\newcommand{\mutox}{\numu\rightarrow\nu_x}
\def\np#1#2#3   {{ Nucl. Phys.} {\bf#1}, #2 (#3)}
\def\pl#1#2#3   {{ Phys. Lett.} {\bf#1}, #2 (#3)}
\def\prev#1#2#3 {{ Phys. Rev.} {\bf#1}, #2 (#3)}
\def\prl#1#2#3 {{ Phys. Rev. Lett.} {\bf#1}, #2 (#3)}
\def\prd#1#2#3 {{ Phys. Rev. D} {\bf#1}, #2 (#3)}
\title{Charged Current Single Pion Cross Section Measurement at MiniBooNE}

\author{M.O. Wascko\address{Department of Physics and Astronomy,\\
                 Louisiana State University, Baton Rouge, LA 70803},
                  for the MiniBooNE Collaboration}
       
\begin{document}

\begin{abstract}
	
We present MiniBooNE's preliminary $\numu \ccpip$ cross section
measurement, calculated using the ratio of $\ccpip$ to CCQE events.
We find the inclusive $\ccpip$ measurement to be below the
\texttt{nuance}~\cite{nuance} and NEUGEN~\cite{neugen} expectations.

\end{abstract}

\maketitle

%------------------------------------------------------------------------
\section{Introduction}

\begin{figure}[t] 
\center
{\includegraphics[width=3.0in]{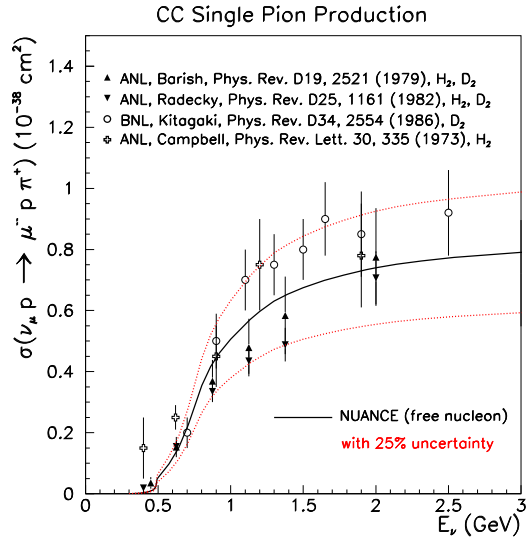}}
\vspace{-0.5in} 
\caption{\em Previous measurements of the total cross section per nucleon 
of the process $\ccpipp$ at low neutrino energy.}
\label{fig:ccpipp} 
\end{figure}

Charged current single pion ($\ccpip$) production has been studied
since the advent of high energy neutrino beams, but the cross section
around 1~GeV energy is still not well understood.  Also, many of the
data that do exist come from hydrogen and deuterium targets, so there
is still a significant need to study nuclear effects in this process.
Figure~\ref{fig:ccpipp}~\cite{ccpipp-plot} shows a comparison of the
$\ccpip$ measurements from the Argonne~\cite{anl} and
Brookhaven~\cite{bnl} bubble chamber experiments, as well as the
\texttt{nuance}~\cite{nuance} Monte Carlo prediction for the $\ccpip$
cross section.

The MiniBooNE $\ccpip$ event sample is a semi-inclusive sample because
the Cherenkov calorimeter detector is not able to resolve the final
state recoil nucleons, and hence cannot distinguish the exclusive
channels.  In this work, we use the label $\ccpip$ to indicate either
of the resonant reactions,
$\nu_{\mu}p\rightarrow\mu^-\Delta^{++}\rightarrow\mu^-p\pi^+$ or
$\nu_{\mu}n\rightarrow\mu^-\Delta^{+}\rightarrow\mu^-n\pi^+$, as well
as coherent pion production, $\ccpipcoh$.  A significant fraction of
the $\ccpip$ events at these energies are expected to arise from
coherent production~\cite{rein-sehgal1}. Given the importance of
$\ccpip$ events as a background for $\mutox$ disappearance
searches~\cite{hiraide}, the recent K2K SciBar limit on CC coherent
pion production at 1.3~GeV~\cite{scibar-coh}, and the importance of
coherent neutral current $\pi^0$ ($\ncpi$) production as a background
source for $\mutoe$ oscillation searches~\cite{hiraide}, there is
significant interest in measuring the $\ccpip$ cross section on
carbon.

MiniBooNE~\cite{boone-prop} is a neutrino oscillation experiment at
Fermilab designed to confirm or rule out the hypothesis that the LSND
$\nuebar$ excess~\cite{lsnd} is due to $\mubartoebar$ oscillations. A
general description of the experiment can be found
elsewhere~\cite{runplan}.  $\ccpip$ events are expected to comprise
$\sim$25\% of the total MiniBooNE neutrino event rate, making these
the second most probable interactions after charged current
quasi-elastic (CCQE).  For this reason they are interesting and useful
for MiniBooNE.  Prior measurements of these cross sections suffered
from poor statistics~\cite{xsec-nu}.
%; the world's published data for these events
%comprises just 7,000 events total, shared among all the experiments
%and over all energies. 
Statistical precision will not be a problem for
MiniBooNE; the MiniBooNE Monte Carlo, which uses the \texttt{nuance}
neutrino generator~\cite{nuance}, predicts that we should have almost
58,000 $\numu \ccpip$ events after cuts in the MiniBooNE detector 
with the full data set of 5.5$\times$10$^{20}$ protons on target (POT).

%------------------------------------------------------------------------
\subsection{Cross Section Ratio Overview \label{sec:overview}}

In this analysis, we normalize the observed rate of $\ccpip$ events to
that of CCQE events, and equate that to the ratio of $\ccpip$ to CCQE
cross sections.  In doing the analysis this way, we use the fact that
the same neutrino flux generates both event samples.  Using the ratio
allows us to neglect the uncertainties in the neutrino flux
prediction.  Moreover, if $\nu_{\mu}$ disappearance were present in
the data, the predicted number of events in the Monte Carlo would be
incorrect because of the depletion of the $\numu$ flux due to $\mutox$
oscillations.  By normalizing to the CCQE data we avoid this issue.

We write the $\ccpip$/CCQE cross section ratio as:
\begin{equation}
\frac{\sigma_{CC1\pi}(E_{\nu})}{\sigma_{CCQE}(E_{\nu})} \ = \ \frac{N^{Data}_{CC1\pi}(E_{\nu})}{N^{Data}_{CCQE}(E_{\nu})},
\label{eqn:xsec}
\end{equation}
where $N^{Data}_{\alpha}$ is the true number of events of type
$\alpha$ in the data.  The true neutrino energy is denoted by
$E_{\nu}$.  If we assume that the CCQE cross section is well simulated
by the Monte Carlo, then we may rewrite Equation \ref{eqn:xsec}:
\begin{equation}
\sigma_{CC1\pi}(E_{\nu}) 
\ = \ 
\frac{N^{Data}_{CC1\pi}(E_{\nu})}{N^{Data}_{CCQE}(E_{\nu})} \times \sigma_{CCQE}^{MC}(E_{\nu}),
\label{eqn:xsec2}
\end{equation}
and thus obtain the $\ccpip$ cross section measurement as a function
of neutrino energy.

The raw event ratio is measured in the Monte Carlo as:
\begin{equation}
  R^{MC}_{raw}(E_{\nu}^{REC}) \ = \ \frac{N^{MC}_{afterCC1\pi cuts}(E_{\nu}^{REC})}{N^{MC}_{afterCCQEcuts}(E_{\nu}^{REC})},
\label{eqn:mc_reweight}
\end{equation}
where we have used $E_{\nu}^{REC}$ to denote the reconstructed energy
of events from either process.  To equate this to the cross section
ratio, we must account for energy smearing, cut efficiencies, and the
presence of background events in each data sample.  We use the Monte
Carlo to estimate each of these factors, and apply the derived
corrections to the data samples before equating the event ratio to the
cross section ratio.

%------------------------------------------------------------------------
\section{$\ccpip$ Event Analysis}

\begin{figure}[t]
\center
{\includegraphics[width=3.0in]{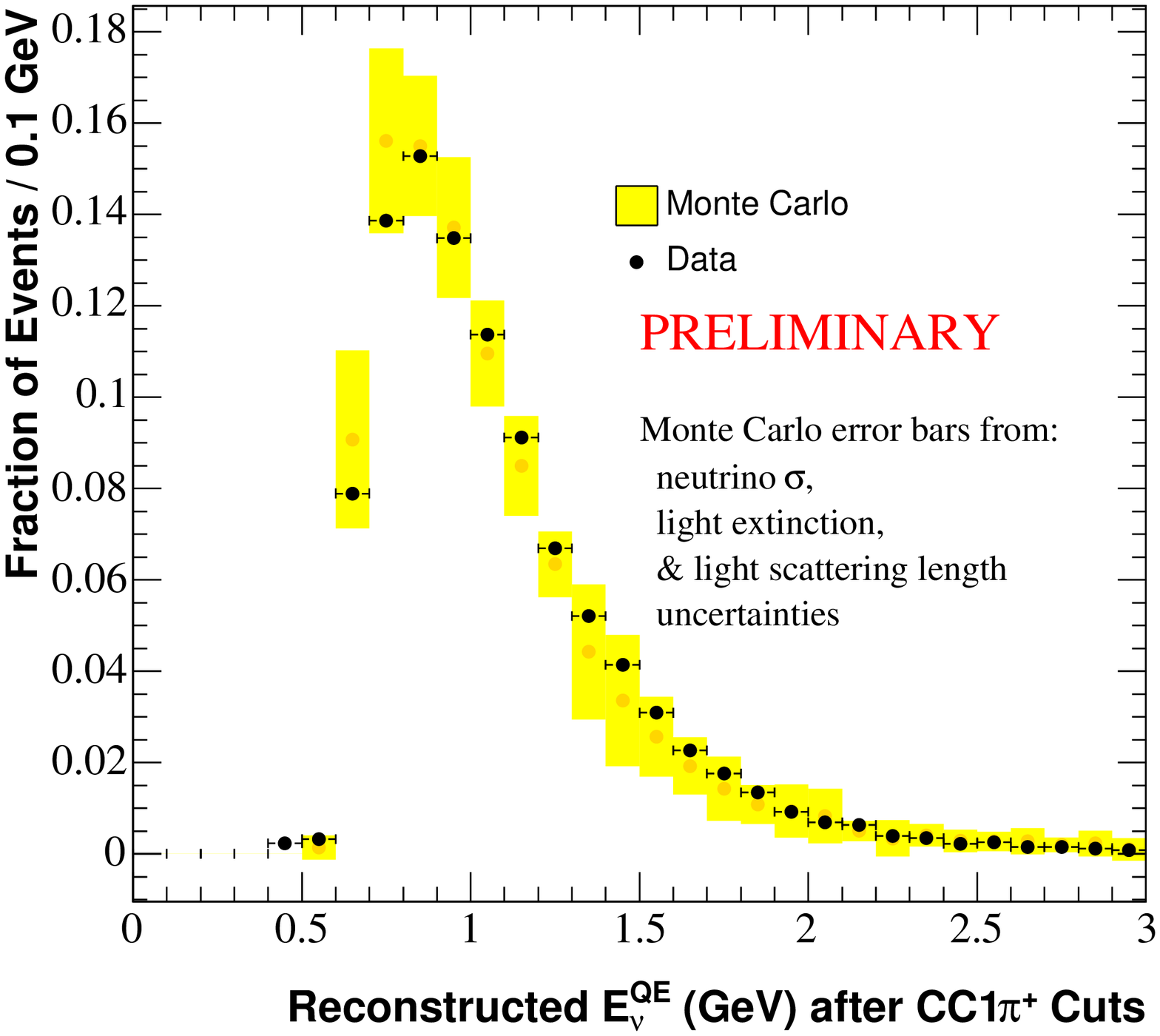}}
\vspace{-0.5in}
\caption{\em Distribution of events versus reconstructed neutrino
energy, for events passing $\ccpip$ cuts.  Black points show data with
statistical errors. Monte Carlo results are shown in yellow, with
error bars showing some important contributions to the systematic
uncertainties.  Data and Monte Carlo are normalized to unit area.}
\label{fig:ccpip_enuqe}
\end{figure}

MiniBooNE's $\ccpip$ event selection requires the simple yet robust
cut of two Michel electrons following the neutrino
interaction~\cite{morgan-ccpip}.  Approximately 40\% of pions emitted
at these energies stop in the detector oil.  These decay to muons
($\mu^+$) with lifetime $\tau=2.6\times10^{-8}$~ns, which then decay
to Michel electrons.  The muons ($\mu^-$) emitted from the neutrino
interaction also come to rest, and 92\% of these decay to Michel
electrons.

\begin{table}[t]
   \caption{\em Event composition of $\ccpip$ sample.}
%, calculated using the \texttt{nuance} Monte Carlo.}
   \label{table:ccpip}
\centering
   \begin{tabular}{ c c }
\\Reaction Type     &  Percentage \\ \hline
  resonant $\ccpip$ &  75.8\%  \\ 
  coherent $\ccpip$ &   9.2\%  \\
  CC QE             &   4.1\%  \\
  multi-pion        &   6.1\%  \\
  DIS               &   2.6\%  \\
  CC$\pi^0$         &   1.5\%  \\
  other             &   0.7\%  \\ \hline
\end{tabular}
\end{table}

Applying this requirement to 3.3$\times$10$^{20}$ POT of MiniBooNE
data yields over 44,000 $\ccpip$ candidate events, making this data
set larger by a factor of five than all previous $\ccpip$ bubble
chamber data published to date.  The Monte Carlo predictions indicate
that the event selection cuts are $\sim$30\% efficient for $\ccpip$
events within the fiducial radius of 500~cm, with a purity of 85\%.
Table~\ref{table:ccpip} shows the fractions of signal and background
events passing the $\ccpip$ event selection.  The background events
come from either events with multiple pions that lose one or more of
them within the nucleus, or CCQE events that acquire a $\pi^+$ though
hadronic interactions within the nucleus, so that the events contain a
single $\mu^-$ and $\pi^+$ in the final state.

The $\ccpip$ events are currently reconstructed under simple
assumptions: we apply a single ring fitter to the PMT hits to find the
position, direction, and photon flux from the Cherenkov ring produced
by the $\mu^-$ in the event.  This yields the muon energy using only
the Cherenkov light in the reconstructed ring; this is done to avoid
including light generated by the $\pi^+$ in calculating the muon
energy.  We then use the fitted energy and direction of the muon to
reconstruct the neutrino energy by assuming the $\ccpip$ reaction is a
simple two body collision, e.g.  $\numu p\rightarrow\mu^-\Delta^{++}$.
Then we write the neutrino energy as:
\begin{equation}
E_{\nu}^{REC} = \frac{1}{2}\frac{ 2m_pE_{\mu} - m_{\mu}^2 + ( m_{\Delta}^2 - m_p^2 )  }{m_p-E_{\mu}+\cos\theta_{\mu}\sqrt{E_{\mu}^2-m_{\mu}^2}},
\label{eq:enu}
\end{equation}
where $m_p$ is the proton mass, $m_{\mu}$ is the muon mass,
$m_{\Delta}$ is the mass of the recoil resonance (which we take to be
exactly 1232~MeV), and $E_{\mu}$ and $\theta_{\mu}$ are the
reconstructed energy and angle of the outgoing muon.

\begin{figure}[t]
\center
{\includegraphics[width=3.0in]{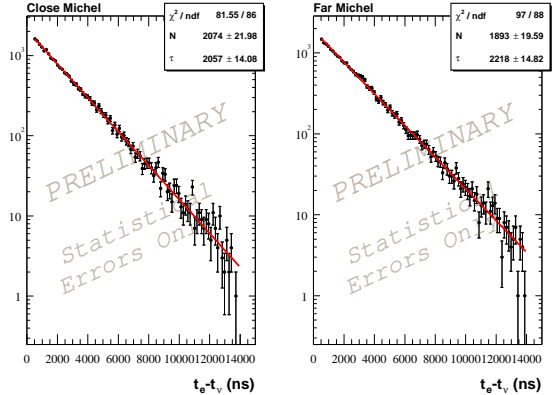}}
\vspace{-0.5in}
\caption{\em Distribution of muon lifetimes from charged current single 
  pion events.  Only data are shown, with only statistical errors.}
\label{fig:ccpip_mu_lifetime}
\end{figure}

Figure~\ref{fig:ccpip_enuqe} shows the fraction of $\ccpip$ events as
a function of reconstructed neutrino energy, for both data and Monte
Carlo.  The data are shown as black points, with error bars
representing the statistical uncertainties.  The Monte Carlo results
are shown as yellow points, with yellow error bands representing the
systematic uncertainties.  Monte Carlo studies indicate that the
neutrino energy resolution is about 20\%.  Much of the resolution
comes from the assumption that the recoil resonance has mass equal to
1232~MeV; however MiniBooNE is not able to reconstruct the invariant
mass of the $\Delta$ because the emitted nucleon does not leave a
Cherenkov ring.  The $\ccpip$ Monte Carlo predictions are based on the
Rein and Sehgal model for resonant and coherent pion
production~\cite{rein-sehgal1}\cite{rein-sehgal2}, and the
\texttt{nuance} nuclear model~\cite{nuance}.

The Michel electrons from the $\ccpip$ candidate events are used to
verify the composition of the data set.  The $\mu^-$ are captured by
carbon nuclei with a probability of 8\%, changing the lifetime to
2026.3$\pm$1.5~ns~\cite{mu+} from the expected
2197.03$\pm$0.04~ns~\cite{mu-}.  The distance from each reconstructed
Michel to the end of the $\mu^-$ track is calculated, the Michels are
sorted into ``close" and ``far" samples based on the distance to the end
of the $\mu^-$ track.  The lifetimes for the two samples are shown in
Fig.~\ref{fig:ccpip_mu_lifetime}. The observed muon lifetimes for the
close and far Michel samples are 2057$\pm$14~ns and 2218$\pm$15~ns,
respectively, indicating that they are indeed mostly from $\mu^-$ and
$\mu^+$ decay, respectively.  The same studies performed on the Monte
Carlo events yield lifetimes of 2043$\pm$15~ns and 2203$\pm$15~ns, and
show that the close and far samples are $\sim$80\% pure $\mu^-$ and
$\mu^+$, repectively.

%------------------------------------------------------------------------
\section{CC QE Event Selection}

\begin{figure}[t]
\center
{\includegraphics[width=3.0in]{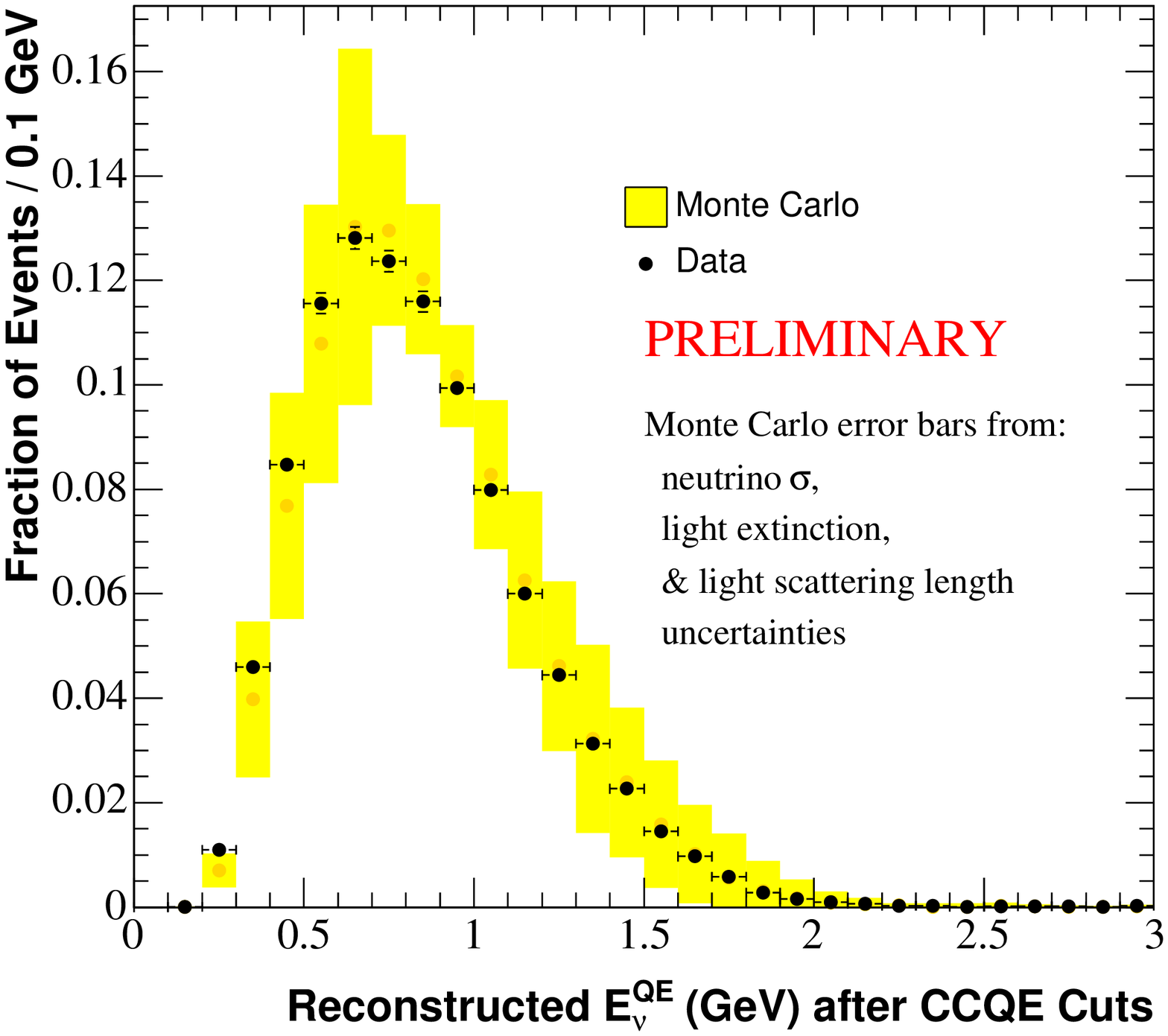}}
\vspace{-0.5in}
\caption{\em Distribution of events versus reconstructed neutrino
energy for events passing CCQE cuts.  Black points show data with
statistical errors.  Monte Carlo results are shown in yellow, with
error bars showing some important contributions to the systematic
uncertainties.  Data and Monte Carlo are normalized to unit area.}
\label{fig:ccqe_enuqe}
\end{figure}

The CCQE event reconstruction is described in greater detail
elsewhere~\cite{jocelyn-ccqe}.  To summarize, the event selection uses
a 10 variable Fisher discriminant designed to find a single Cherenkov
ring from a muon. The neutrino energy is reconstructed using simple
two-body kinematics and corrected for the effects of Fermi momentum of
the target nucleon.  Figure~\ref{fig:ccqe_enuqe} shows the fraction of
CCQE events as a function of reconstructed neutrino energy, for both
data and Monte Carlo.  The data are shown as black points, with error
bars representing the statistical uncertainties.  The Monte Carlo
results are shown as yellow points, with yellow error bands
representing the systematic uncertainties.

Applying the CCQE event selection to 2.3$\times$10$^{20}$ POT of
MiniBooNE data yields over 60k CCQE events, with 86\% purity.
Table~\ref{table:ccqe} shows the fractions of signal and background
events passing the CCQE event selection.  The background events come
mostly from single pion events in which the pion is absorbed within
the nucleus, so that the events contain a single $\mu^-$ and no $\pi$
in the final state.

Monte Carlo studies indicate that the neutrino energy resolution is
about 10\%.  This is better than the $\ccpip$ energy resolution
because the mass of the recoil particle (a proton) is well known, and
the muon Cherenkov ring is cleaner because there is no $\pi^+$ in
the final state.

\begin{table}[t]
   \caption{\em Event composition of CCQE sample.}
%, calculated using the \texttt{nuance} Monte Carlo.}
   \label{table:ccqe}
\centering
   \begin{tabular}{ c c }
\\Reaction Type      &  Percentage \\ \hline
  CCQE               &  86.0\%  \\ 
  $\ccpip$: resonant &   8.9\%  \\
  $\ccpip$: coherent &   1.4\%  \\
  NC1$\pi^+$         &   1.7\%  \\
  CC1$\pi^0$         &   1.1\%  \\
  CC$\pi^0$          &   1.5\%  \\
  other              &   0.9\%  \\ \hline
\end{tabular}
\end{table}

%

%------------------------------------------------------------------------
\section{$\ccpip$ Cross Section}

\begin{figure}[t]
\center
{\includegraphics[width=3.0in]{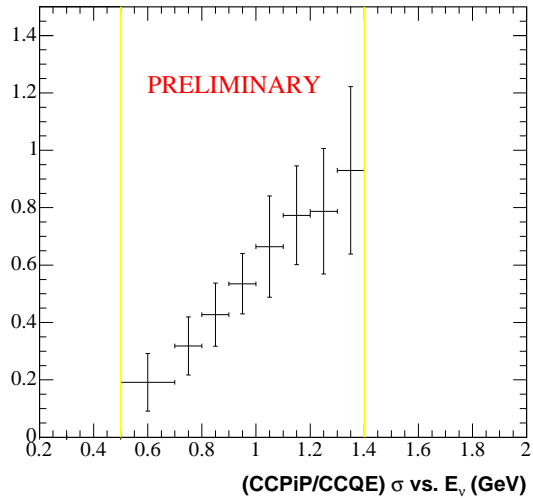}}
\vspace{-0.5in}
\caption{\em Ratio of $\ccpip$/CCQE events vs. neutrino energy, after
corrections for background events, energy smearing, and cut
efficiencies.  The black error bars show the statistical and
systematic uncertainties added in quadrature.}
\label{fig:ccpip_ratio}
\end{figure}
To use Equation~\ref{eqn:xsec2} to calculate the cross section, we
first calculate the ratio of $\ccpip$/CCQE events.  The numerator of
the ratio is the event sample shown in Figure~\ref{fig:ccpip_enuqe},
after it was corrected for the background fraction, energy smearing
and cut efficiencies.  Similarly, the denominator is shown in
Figure~\ref{fig:ccqe_enuqe}.  The ratio of $\ccpip$/CCQE events as a
function of neutrino energy is shown in Figure~\ref{fig:ccpip_ratio}.
Note that this neutrino energy has been corrected for energy smearing.
The error bars show the systematic uncertainties added in quadrature
with the statistical uncertainties.

We restrict the analysis to the reconstructed energy region
0.5~GeV-1.4~GeV.  This is because the $\ccpip$ sample has insufficient
statistics below $\sim$0.5~GeV, due to the energy threshold for
$\Delta$ production, and the CCQE sample has insufficient statistics
above $\sim$1.4~GeV, due to the efficiency of the event selection
cuts for high muon energies.

To calculate the $\ccpip$ cross section, we multiply the ratio by the
Monte Carlo CCQE prediction.  The Monte Carlo prediction is based on
the \texttt{nuance} neutrino generator, which assumes a
Llewellyn-Smith quasi-elastic cross section~\cite{llewelyn-smith}, and
a Smith-Moniz Fermi gas model for nuclear
interactions~\cite{smith-moniz}.  The \texttt{nuance} program also
simulates final state interactions, which mainly include the
interaction probabilities of the final state particles while they are
still inside the nucleus~\cite{nuance}.  The MiniBooNE Monte Carlo
uses the BBA03 non-dipole vector form factors~\cite{bba03}, and a
dipole axial form factor with M$_A$=1.03.

\begin{figure}[t]
\center
{\includegraphics[width=3.0in]{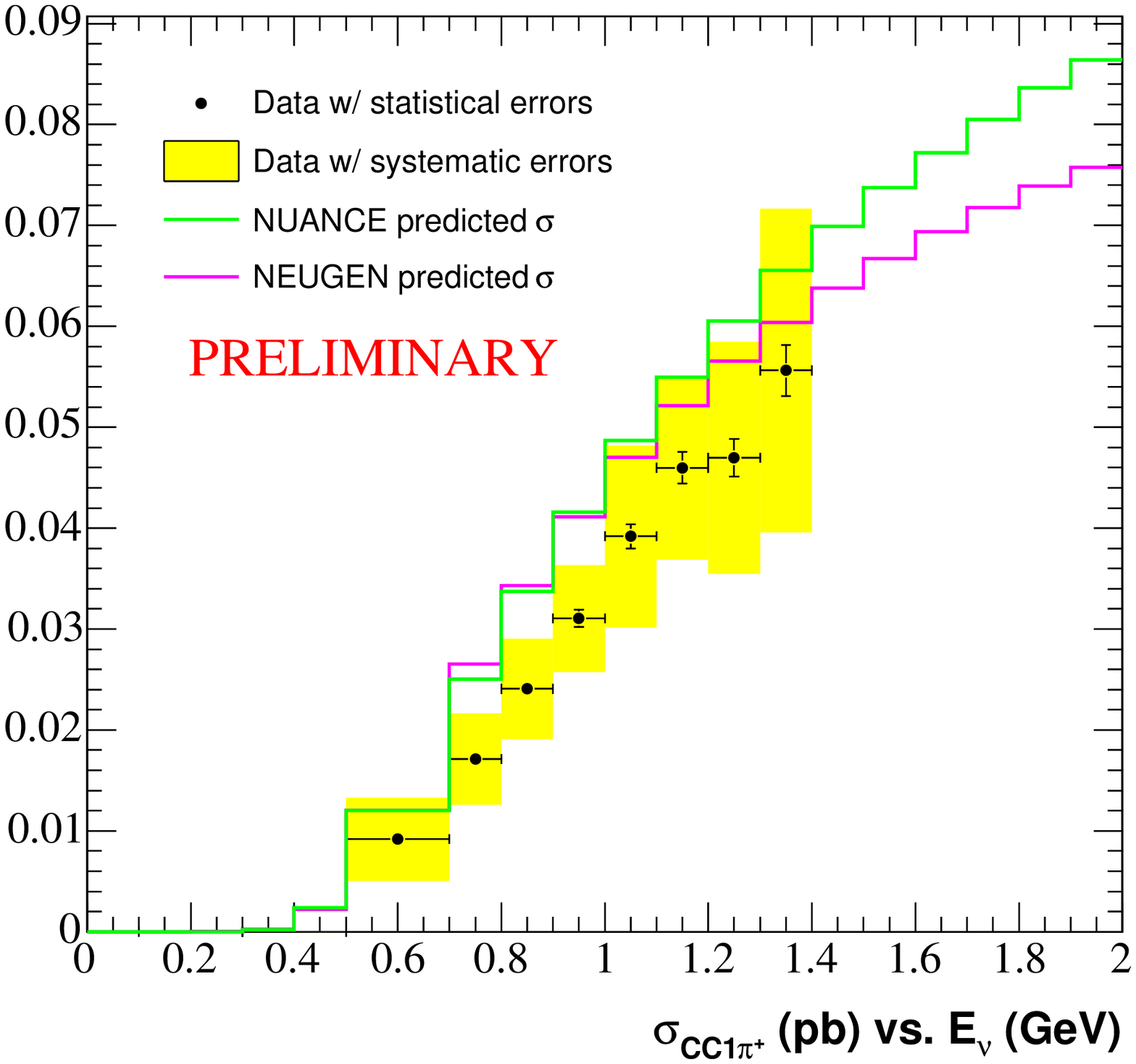}}
\vspace{-0.5in}
\caption{\em Semi-inclusive $\ccpip$ cross section vs. neutrino
energy, extracted assuming the \texttt{nuance} CCQE cross section
prediction with M$_A^{QE}$=1.03~GeV.  Statistical uncertainties are
shown by the black error bars, and the important contributions to 
the systematic uncertainties are shown by the
yellow error bands.  Also shown are the \texttt{nuance} and NEUGEN
predictions for the semi-inclusive $\ccpip$ cross section vs. 
reconstructed neutrino energy.}
\label{fig:ccpip_xsec}
\end{figure}

Figure~\ref{fig:ccpip_xsec} shows MiniBooNE's $\ccpip$ cross section
measurement.  Again, the data are shown as black points, with
statistical error bars.  The yellow error bands represent the present
estimate of the systematic uncertainties.  Also shown are the
\texttt{nuance} and NEUGEN~\cite{neugen} predictions for the $\ccpip$
cross section.  Although the $\ccpip$ measurement is calculated from
the ratio with the \texttt{nuance} CCQE cross section, the comparison
to NEUGEN is appropriate because its CCQE prediction is identical to
the \texttt{nuance} prediction.

Note that the MiniBooNE $\ccpip$ measurement lies $\sim$25\% below the
\texttt{nuance} and NEUGEN predictions.  One plausible explanation for
this apparent disparity is offered by Figure~\ref{fig:ccpipp}: the
ANL\cite{anl} and BNL\cite{bnl} measurements of the $\ccpipp$ cross
section.  The BNL measurement lies $\sim$40\% above the ANL, and
\texttt{nuance} splits the difference.  MiniBooNE's measurement lying
25\% below the \texttt{nuance} prediction suggests that it is more
consistent with the ANL observation.

%------------------------------------------------------------------------
\section{Systematic Uncertainties}

The sources of systematic error considered here are summarized in
Table \ref{sys_source_table}, and can be generally categorized as
neutrino cross section, optical model, and energy scale uncertainties.
By optical model, we mean the properties of light generation and
transmission in the detector mineral oil.  The systematic
uncertainties are assessed on the $\ccpip$/CCQE ratio, so there is
some cancellation of errors.

We include cross section uncertainties in a cross section measurement
because we must assess the error on the predicted number of events
which pass the CCQE selection cuts.  The CCQE background after cuts is
dominantly from $\ccpip$ events, and so any normalization uncertainty
on the predicted $\ccpip$ (and CCQE) event rates must be included.
Second, efficiency corrections are applied to the $\ccpip$ data which
are derived from the Monte Carlo, and so any energy dependent
uncertainties must be included.

The cross section uncertainties are all derived from external
data~\cite{hawker}.  The optical model uncertainties are derived from
``table-top" measurements of the oil and {\em in situ} calibration
data samples~\cite{optical-model}.  The correlations between the
sources of uncertainty are not yet fully determined, so we present the
analysis assuming they are all uncorrelated, adding all uncertainties
in quadrature.  In fact, we expect many of the sources of uncertainty
to be {\em anti-}correlated.

\begin{table}
\caption{ \label{sys_source_table} Sources of uncertainty considered
for the $\ccpip$ cross section analysis, summed over the entire
analysis region from 0.5 to 1.4~GeV.}
\begin{center}
\begin{tabular}{l c}
\emph{source} & {\em effect on $\sigma$}\\
\hline
Optical Model  & 20\%\\
Cross Sections & 15\% \\
Energy Scale   & 10\%\\
\hline
Statistics     &  6\%\\
\hline
\end{tabular}
\end{center}
\end{table}

%------------------------------------------------------------------------
\section{Conclusions}

We have presented the first measurement of the semi-inclusive $\numu
\ccpip$ cross section ratio on a nuclear target near 1~GeV.  From
this, we extract a $\ccpip$ cross section measurement which is lower
than the predictions of the \texttt{nuance} and NEUGEN Monte Carlos,
but appears consistent with the ANL observation of resonant single
pion production.

%In addition to the present measurements, MiniBooNE is developing a
%method to extract the fraction of coherent events in the $\ccpip$
%sample, as well as a $\mutoe$ oscillation search in the $\ccpip$
%channel.

%------------------------------------------------------------------------
\section{Acknowledgments}

The author is pleased to acknowledge collaboration with J. Monroe in
this analysis, and would also like to express gratitude to the
organizers of NuInt05 for their generous travel support.

The MiniBooNE collaboration gratefully acknowledges support from the
Department of Energy and the National Science Foundation. The author
was supported by grant number DE-FG02-91ER0617 from the Department of
Energy.


\begin{thebibliography}{9}

\bibitem{nuance} D.~Casper, Nucl. Phys. Proc. Suppl. {\bf 112}, 161
                 (2002), {\em hep-ph/0208030}.

\bibitem{neugen} H.~Gallagher, Nucl. Phys. Proc. Suppl. {\bf 112} (2002) 188.

\bibitem{ccpipp-plot} G.~P.~Zeller, private communication.

\bibitem{anl} S.~J.~Barish {\em et al.}, \prd {19}{2521}{1979}, G.~M.~Radecky {\em et al.} \prd {25}{1161}{1982}.

\bibitem{bnl} T.~Kitagaki {\em et al.}, \prd {34}{2554}{1986} .


\bibitem{rein-sehgal1} D.~Rein and L.~M.~Sehgal, Ann. Phys. 133 (1981) 79 .

\bibitem{hiraide} K.~Hiraide, ``The SciBooNE Experiment,'' these proceedings.

\bibitem{scibar-coh} M.Hasegawa {\em et al.}, \prl{95}{2005}{252301} .

\bibitem{boone-prop} FERMILAB-PROPOSAL-0898, Dec 1997, {\em nucl-ex/9706011} .

\bibitem{lsnd} A. Aguilar {\em et al.},Phys. Rev. D, 64, 2001, {\em hep-ex/0104049}.

\bibitem{runplan} A.~A.~Aguilar-Arevalo {\em et al.}, ``The MiniBooNE
  Run Plan'' (2003).

\bibitem{xsec-nu} G.~P.~Zeller, NuInt02,  {\em hep-ex/0312061}.

\bibitem{morgan-ccpip} M.~O.~Wascko, DPF04,  {\em hep-ex/0412008}.

\bibitem{rein-sehgal2} D.~Rein and L.~M.~Sehgal, \np{B223}{29}{1983} .

\bibitem{mu+} S.Eidelman {\em et al.}, \pl{B592}{33}{2004} .

\bibitem{mu-} T.Suzuki {\em et al.}, \prev{C35}{2122}{1987} .

\bibitem{jocelyn-ccqe} J.~Monroe,  Moriond 2004, {\em hep-ex/0406048}.

\bibitem{llewelyn-smith} C.~Llewellyn-Smith, Phys. Rept. C3 (1972) 261 .

\bibitem{smith-moniz} R.~A.~Smith and E.~J.~Moniz, Nucl. Phys. B 43 (1972) 605. [Erratum-ibid. B 101 (1975) 547].

\bibitem{bba03} H.~Budd {\em et al.}, {\em hep-ex/0308005}, {\em hep-ex/0309024} .

\bibitem{hawker} E.~A.~Hawker, Nucl. Phys. Proc. Suppl. 139, 260 (2005).

\bibitem{optical-model} B.~C.~Brown, FERMILAB-CONF-04-282-E

\end{thebibliography}
\end{document}